\documentclass[showpacs,preprintnumbers,amsmath,amssymb,12pt,floatfix,epsfig]{revtex4}
\usepackage{graphicx}
\begin{document}
\draft
\title{Neutral-current Neutrino-nucleus Scattering in Quasielastic Region }
\author{K. S. Kim$^{1)}$\thanks{kyungsik@hau.ac.kr}, B. G. Yu${^1)}$,
M. K. Cheoun${^2)}$, T. K. Choi${^3)}$, and M. T. Chung${^4)}$}
\address{1)School of Liberal Arts and Science, Korea Aerospace University,
Koyang 200-1, Korea  \\
2)Department of Physics, Soongsil University, Seoul, 156-743, Korea \\
3)Department of Physics, Yonsei University, Wonju, 220-710 Korea \\
4)Department of Radiation, Dongsin University, Naju, 520-714 Korea
}

\begin{abstract}
The neutral-current neutrino-nucleus scattering is calculated
through the neutrino-induced knocked-out nucleon process in the
quasielastic region by using a relativistic single particle model
for the bound and continuum states. The incident energy range
between 500 MeV and 1.0 GeV is used for the neutrino
(antineutrino) scattering on $^{12}$C target nucleus. The effects
of the final state interaction of the knocked-out nucleon are
studied not only on the cross section but also on the asymmetry
due to the difference between neutrinos and antineutrinos, within
a relativistic optical potential. We also investigate the
sensitivity of the strange quark contents in the nucleon on the
asymmetry.
\end{abstract}
\narrowtext
\maketitle

\section{Introduction}
Neutrino-nucleus scattering has become to be widely interested in
different fields of physics such as astrophysics, cosmology,
particle, and nuclear physics. Such interests are found not only
in the nuclear astrophysics but also in the nuclear physics
itself. In particular, the neutral-current scattering of neutrinos
and antineutrinos on nuclei is used to determine the structure of
hadronic weak neutral currents. Along this line, Brookhaven
National Laboratory (BNL) \cite{brook} reported that the value of
a strange axial vector form factor of the nucleon does not have
zero. The primary goal of the BooNE experiment \cite{boone} was to
search for neutrino oscillation to test neutrino mass and then
detected the first anti-neutrino events in January of 2006. The
CNGS project \cite{cngs} was proposed to detect the $\tau$
production through $(\nu_{\tau}, \tau^-)$ or $({\bar
{\nu_{\tau}}}, \tau^+)$ reactions, which will send a neutrino beam
from CERN to the Gran Sasso laboratory.

At intermediate energies, there are many theoretical works
\cite{garvey,hung,umino,alberico,jacowicz,giusti1,giusti2,udias}
for the neutrino-nucleus scattering. The relativistic Fermi gas
(RFG) model in Refs. \cite{hung,umino} was applied to study the
contribution of the strange quark to the nucleon form factor in
the neutrino-induced knocked-out nucleon process. In Ref.
\cite{alberico}, the relativistic plane wave impulse approximation
(RPWIA) calculations were compared with the RFG calculations using
a relativistic shell model. In particular, Ref. \cite{jacowicz}
suggested a method of identifying neutrinos and antineutrinos by
estimating the polarization asymmetry stemming from the
differences of the intrinsic helicities of them. The calculation
is carried out within a nonrelativistic nuclear shell model under
a Woods-Saxon potential for the bound states.

For the final state interaction (FSI) of the knocked-out nucleon,
the authors in Ref. \cite{giusti1} showed that there is no flux
loss to take into account the complex optical potential in the
charged-current reaction. In Ref. \cite{giusti2}, the importance
of the FSI and the contribution of the strange quark content were
shown in the neutral-current reaction by using a relativistic
optical potential. Ref. \cite{udias} presented the comparison of
the relativistic distorted wave impulse approximation (RDWIA) by
Madrid group \cite{madrid} with the relativistic multiple
scattering Glauber approximation developed by Ghent group
\cite{ghent}.

On the other hand, Ohio University group \cite{jin1,jin2}
calculated inclusive $(e,e')$ and exclusive $(e,e'p)$ reactions in
quasielastic region using partial wave expansions of the electron
wave functions in the distorted wave Born approximation (DWBA)
under the presence of the electron Coulomb distortion due to the
target nucleus. However, the DWBA calculations do not allow a
separation of the cross section into a longitudinal part and a
transverse part and are numerically challenging, and computational
time increases rapidly with higher incident electron energies. In
order to avoid such difficulties of the DWBA calculations, Kim and
Wright \cite{kim1,kim2} developed an approximate treatment of the
electron Coulomb distortion which does allow the separation of the
cross section into a longitudinal part and a transverse part.

In addition, Ref. \cite{kim1} showed a very good description of
quasielastic scattering processes using a relativistic single
particle model which requires the wave functions of bound and
continuum nucleons and a transition current operator. The bound
state wave functions are obtained from solving the Dirac equation
in the presence of the strong vector and scalar
potentials\cite{horo}. For the inclusive $(e,e')$ reaction where
the knocked-out nucleons are not observed, the continuum wave
functions are solutions to a real potential so as not to lose any
flux. This ansatz guarantees the current conservation and gauge
invariance. For the exclusive $(e,e'p)$ reaction, Kim and Wright
\cite{kim2} used the relativistic optical potential to solve the
wave function of the knocked-out proton, which is generated by
Ohio State University group \cite{clark}. These theoretical
results explained experimental data very well with no free
parameters except an overall scale factor, called the
spectroscopic factor.

In this paper, we present the neutral-current neutrino-nucleus
scattering in the quasielastic region, where the inelastic
processes like pion production and delta resonance are excluded.
The incident neutrino (antineutrino) energies are concerned in
intermediate ranges (between 500 MeV and 1.0 GeV). The
relativistic bound state wave functions are obtained from solving
a Dirac equation in the presence of strong scalar and vector
potentials based on the $\sigma - \omega$ model \cite{horo}. In
order to investigate the effect of the FSI we take account into
the relativistic optical potential \cite{clark}. This nuclear
model is the same model as used in Ref. \cite{kim2} except the
electron Coulomb distortion. Furthermore, the effects of the
strangeness in the axial form factor of the weak current operator
are also studied.

The outline of this paper is as follows. In Sec. II we address the
formalism and the numerical results are shown in Sec. III. Our
summary of the results and conclusions are given in Section IV.

\section{Formalism}
The neutrino-nucleus scattering is described by the connection of
the electromagnetic interaction and the weak interaction. The
four-momenta of the incident and outgoing neutrinos
(antineutrinos) are labelled $p_i^{\mu}=(E_i, {\bf p}_i)$ and
$p_f^{\mu}=(E_f, {\bf p}_f)$. We choose the nucleus fixed frame
where the target nucleus is seated at the origin of the coordinate
system. $p_A^{\mu}=(E_A, {\bf p}_A)$, $p_{A-1}^{\mu}=(E_{A-1},
{\bf p}_{A-1})$, and $p^{\mu}=(E_p, {\bf p})$ represent the
four-momenta of the target nucleus, the residual nucleus, and the
knocked-out nucleon, respectively. In the laboratory frame, the
inclusive cross section is given by the contraction between the
lepton tensor and the hadron tensor \cite{udias}
\begin{eqnarray}
{\frac {d\sigma} {dE_f}} = 4\pi^2{\frac {M_N M_{A-1}} {(2\pi)^3
M_A}} \int \sin \theta_l d\theta_l \int \sin \theta_p d\theta_p p
f^{-1}_{rec} \sigma^Z_M [v_L R_L + v_T R_T + h v'_T R'_T ],
\label{cs}
\end{eqnarray}
where $M_N$ is the nucleon mass, $\theta_l$ denotes the scattering
angle of the lepton, and $h=-1$ $(h=+1)$ corresponds to the
helicity of the incident neutrino (antineutrino). The squared
four-momentum transfer is given by $Q^2=q^2 - \omega^2$.
$\sigma^Z_M$ is defined by
\begin{equation}
\sigma^Z_M = \left ( {\frac {G_F \cos (\theta_l/2) E_f M_Z^2}
{{\sqrt 2} \pi (Q^2 + M^2_Z)}} \right ),
\end{equation}
where $G_F$ is the Fermi constant given by $G_F \simeq 1.16639
\times 10^{-11}$ MeV$^{-1}$ and $M_Z$ is the rest mass of
$Z$-boson. The recoil factor $f_{rec}$ is given by
\begin{equation}
f_{rec} = {\frac {E_{A-1}} {M_A}} \left | 1 + {\frac {E_p}
{E_{A-1}}} \left [ 1 - {\frac {{\bf q} \cdot {\bf p}} {p^2}}
\right ] \right |. \end{equation}

For the neutral-current reaction, the coefficients $v$ are given
by
\begin{equation}
v_L=1, \;\;\;\;\;\;\;  v_T=\tan^2 {\frac {\theta_l} {2}} + {\frac
{Q^2} {2q^2}}, \;\;\;\;\;\;\; v'_T=\tan {\frac {\theta_l} {2}}
\left [\tan^2 {\frac {\theta_l} {2}} + {\frac {Q^2} {q^2}} \right
]^{1/2}.
\end{equation}
The corresponding response functions are given by
\begin{eqnarray}
R_L=\left | J^0 - {\frac {\omega} {q}} J^z \right |^2,
\;\;\;\;\;\;\; R_T=|J^x|^2 + |J^y|^2, \;\;\;\;\;\;\; R'_T = 2
{\mbox {Im}}({J^x}^*J^y).
\end{eqnarray}

The nucleon current $J$ represents the Fourier transform of the
nucleon current density written as
\begin{equation}
J^{\mu}=\int {\bar \psi}_p {\hat {\bf J}}^{\mu} \psi_b e^{i{\bf
q}{\cdot}{\bf r}}d^3r,
\end{equation}
where ${\hat {\bf J}}^{\mu}$ is a free nucleon current operator,
and $\psi_{p}$ and $\psi_{b}$ are the wave functions of the
knocked-out and the bound state nucleons, respectively. For a free
nucleon, the current operator comprises the weak vector and the
axial form factors related to the eletromagnetic current given by
\begin{equation}
{\hat {\bf J}}^{\mu}=F_{1}^V (Q^2){\gamma}^{\mu}+ F_{2}^V
(Q^2){\frac {i \kappa} {2M_N}}{\sigma}^{\mu\nu}q_{\nu} + G_A(Q^2)
\gamma^{\mu} \gamma^5 + {\frac {1} {2M_N}}G_P(Q^2) q^{\mu}
\gamma^5,
\end{equation}
where $\kappa$ is the anomalous magnetic moment. The weak vector
form factors for protons ($F_{i}^{V,p} (Q^2)$) and neutrons
($F_{i}^{V,n} (Q^2)$), by the conservation of the vector current
(CVC) hypothesis with inclusion of an isoscalar strange quark
($F_i^s$) contribution, are given by \cite{giusti2}
\begin{eqnarray}
F_i^{V,~ p(n)}&=&({\frac 1 2} - 2 \sin^2 \theta_W ) F_i^{p(n)} -
{\frac 1 2} F_i^{n(p)} -{\frac 1 2} F_i^s,
\end{eqnarray}
where $\theta_W$ is the Weinberg angle given by $\sin^2 \theta_W =
0.2224$.

The strange vector form factors are expressed as \cite{garvey}
\begin{equation}
F_1^s(Q^2) = {\frac {F_1^s Q^2} {(1+\tau)(1+Q^2/M_V^2)^2}},
\;\;\;\;\; F_2^s(Q^2) = {\frac {F_2^s(0)}
{(1+\tau)(1+Q^2/M_V^2)^2}},
\end{equation}
where $\tau=Q^2/(4M_N^2)$, $M_V=0.843$ GeV,
$F_1^s=-<r_s^2>/6=0.53$ GeV$^{-2}$, and $F_2^s(0)=\mu_s$ is a
strange magnetic moment given by $\mu_s=-0.4$.

The axial form factor is given by \cite{musolf}
\begin{eqnarray}
G_A&=&{\frac 1 2} (\mp g_A + g_A^s)G, \label{gs}
\end{eqnarray}
where $g_A=1.262$, $G=1/(1+Q^2/M^2)^2$ with $M=1.032$ GeV, and
$g_A^s=-0.19$, which repersents the strange quark contents on the
nucleon \cite{bernard}. $-(+)$ coming from the isospin dependence
denotes the knocked-out proton (neutron), respectively.

The induced pseudoscalar form factor from the Goldberger-Treiman
relation is parametrized as
\begin{equation}
G_P(Q^2) = {\frac {2M_N} {Q^2+m^2_{\pi}}} G_A,
\end{equation}
where $m_{\pi}$ is the pion mass. The contribution of the
pseudoscalar form factor vanishes for the neutral-current reaction
because of the final lepton mass participating in this reaction.

\section{results}
We calculate the neutral-current neutrino-nucleus scattering for
$^{12}$C nucleus in the quasielastic region, where the inelastic
processes like pion production, Delta resonance, and etc are
excluded. Within the relativistic framework, the wave functions of
the bound state are generated by the $\sigma - \omega$ model
\cite{horo} and the continuum states of the knocked-out nucleon
are obtained by solving a Dirac equation in the presence of the
phenomenological relativistic optical potentials generated by
EDAD1 \cite{clark}. We use two incident neutrino (antineutrino)
energies, 500 MeV and 1.0 GeV, where the inelastic contributions
are as small as negligible.

\begin{figure}
\includegraphics[width=0.5\linewidth]{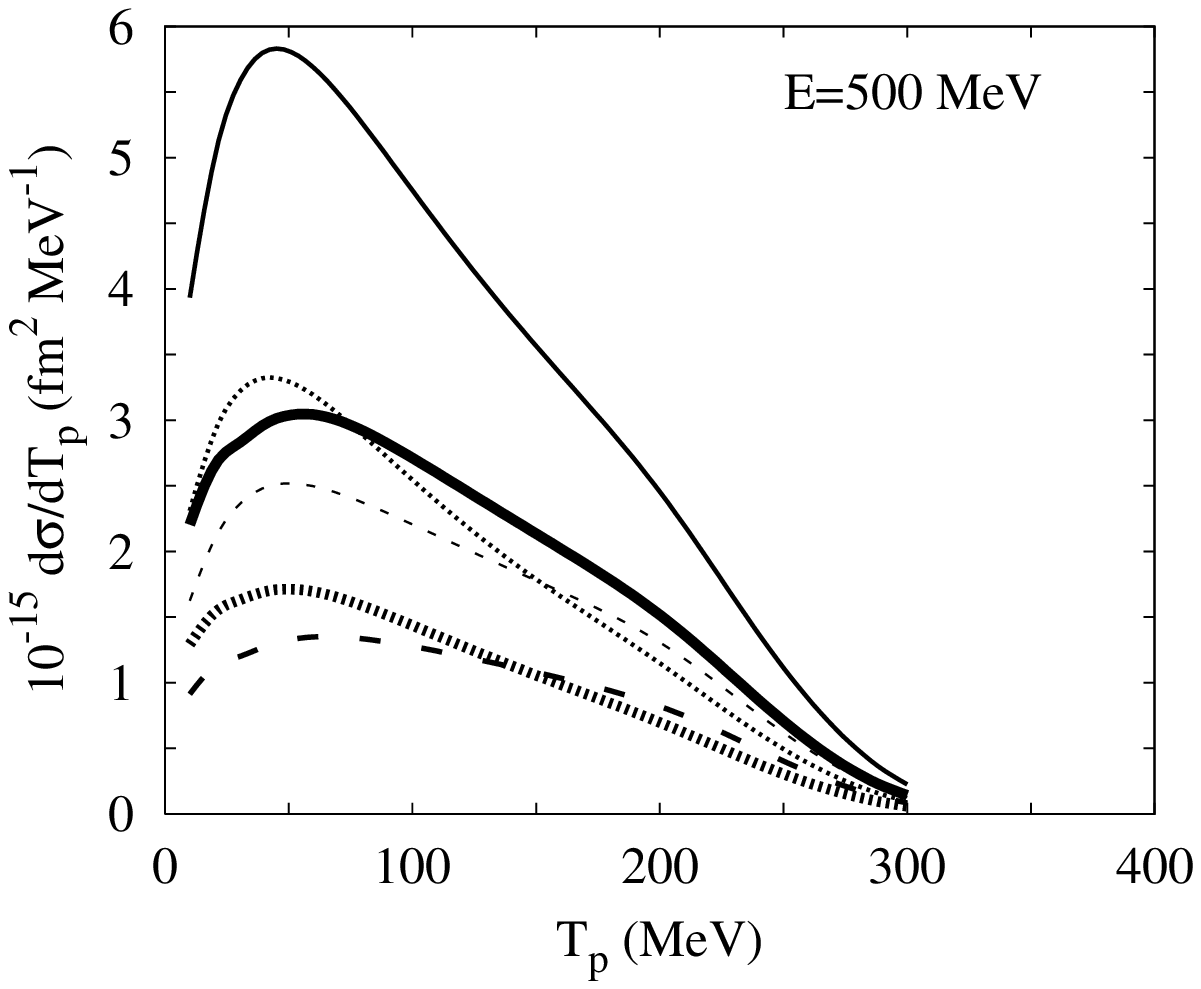}
\includegraphics[width=0.5\linewidth]{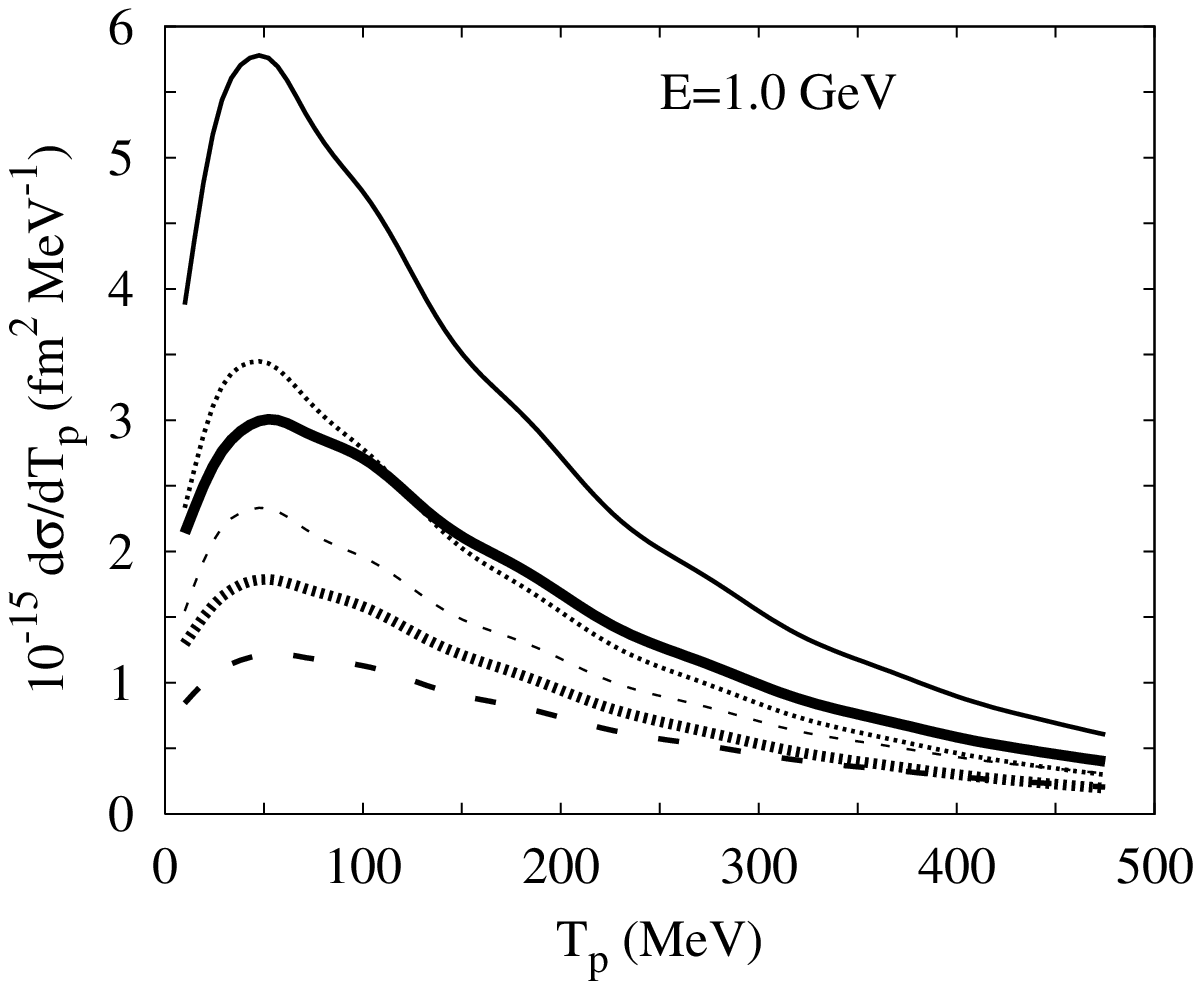}
\caption{Neutral current $^{12}$C($\nu, \nu'$) cross sections as a
function of the knocked out nucleon kinetic energy $T_p$ at
incident neutrino energies $E=500$ MeV and 1.0 GeV. Solid curves
are the results for the cross sections, dashed and dotted lines
are the contributions of the proton and the neutron, respectively.
Thick and thin lines are calculations with the relativistic
optical potential and with no any potential of the knocked-out
nucleon.} \label{neut}
\end{figure}

\begin{figure}
\includegraphics[width=0.5\linewidth]{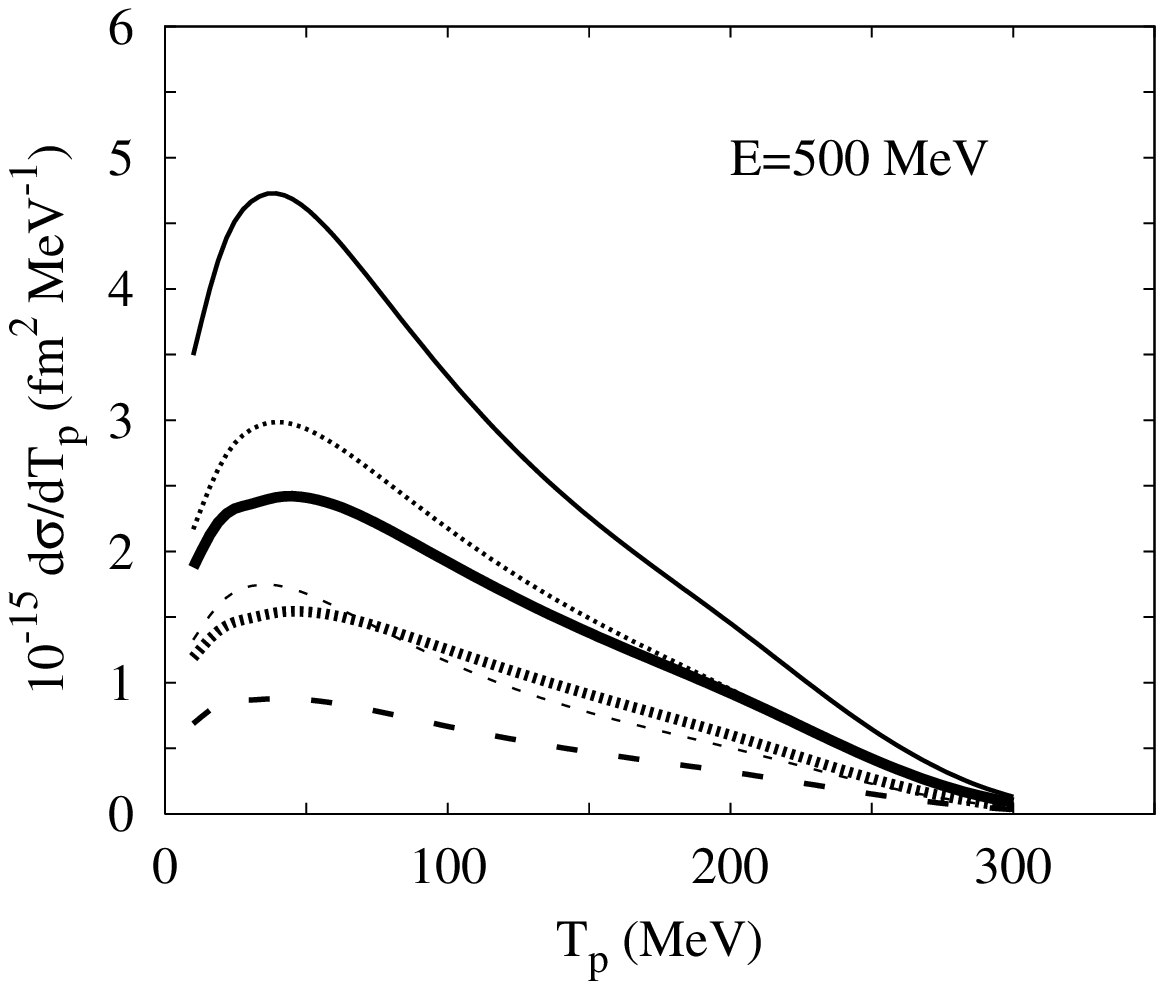}
\includegraphics[width=0.5\linewidth]{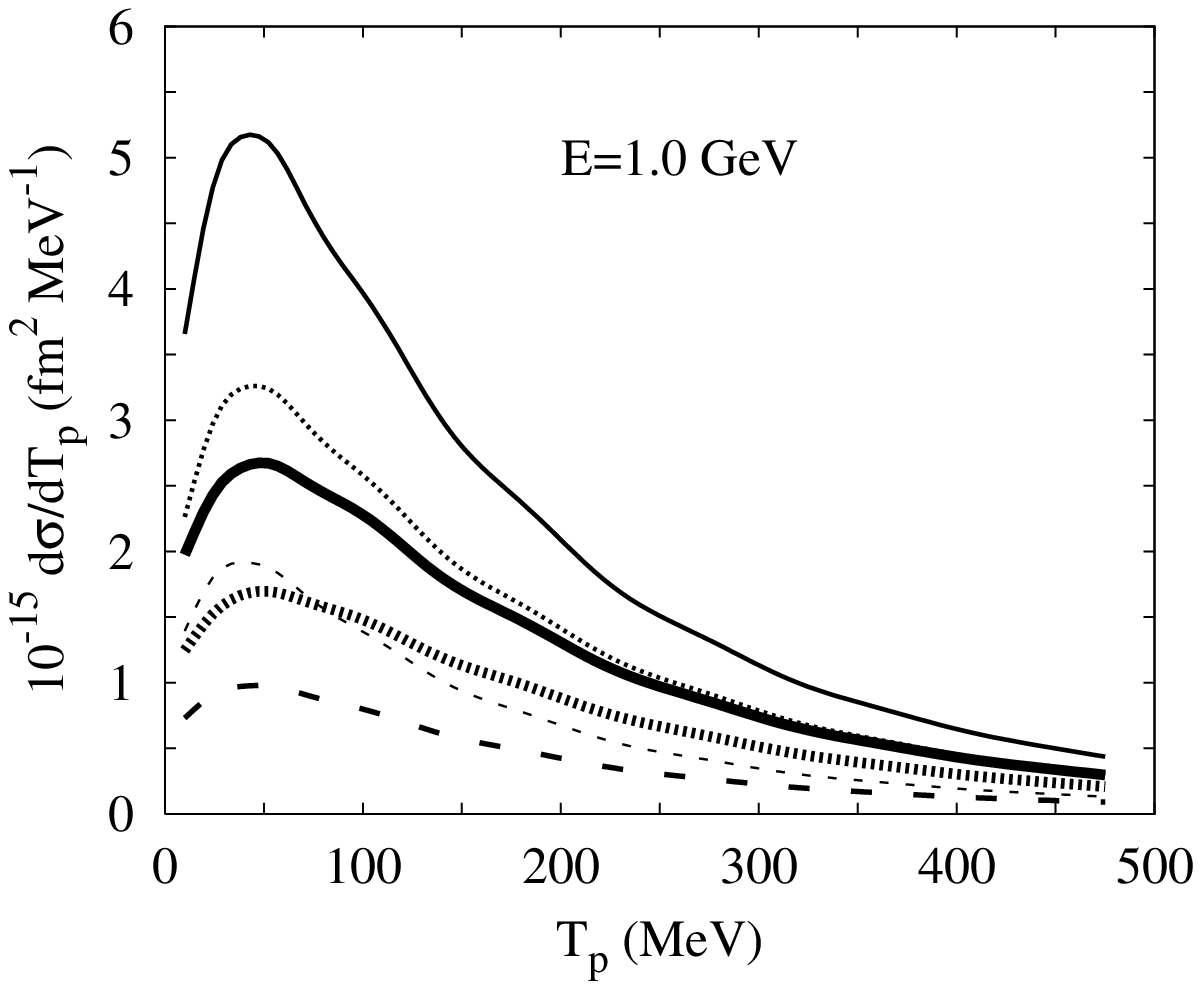}
\caption{The same as in Fig. \ref{neut} but for the antineutrino.}
\label{anti}
\end{figure}

In Figs. \ref{neut} and \ref{anti}, we show the inclusive cross
sections for the neutrino-nucleus and antineutrino-nucleus
scattering of the neutral-current reactions as a function of the
knocked-out nucleon kinetic energy $T_p$ at incident neutrino
energies $E=500$ MeV and 1.0 GeV. Solid curves present the results
for the cross sections, and dashed and dotted lines show the
contributions of the knocked-out protons and the knocked-out
neutrons, respectively. Thick (thin) curves are the results {\bf
with (without)} the optical potential. The effects of the FSI
produce a reduction of the cross section around 50\%, which is the
same result as Refs. \cite{giusti2,udias}. Therefore the FSI of
the knocked-out nucleon is confirmed again to be a vital
ingredient in the neutrino-nucleus reactions. The knocked-out
neutrons contribute to the cross section larger than the
knocked-out protons. The ratios of the contribution between the
knocked-out protons and neutrons are similar with/without the
optical potential.

\begin{figure}
\includegraphics[width=0.5\linewidth]{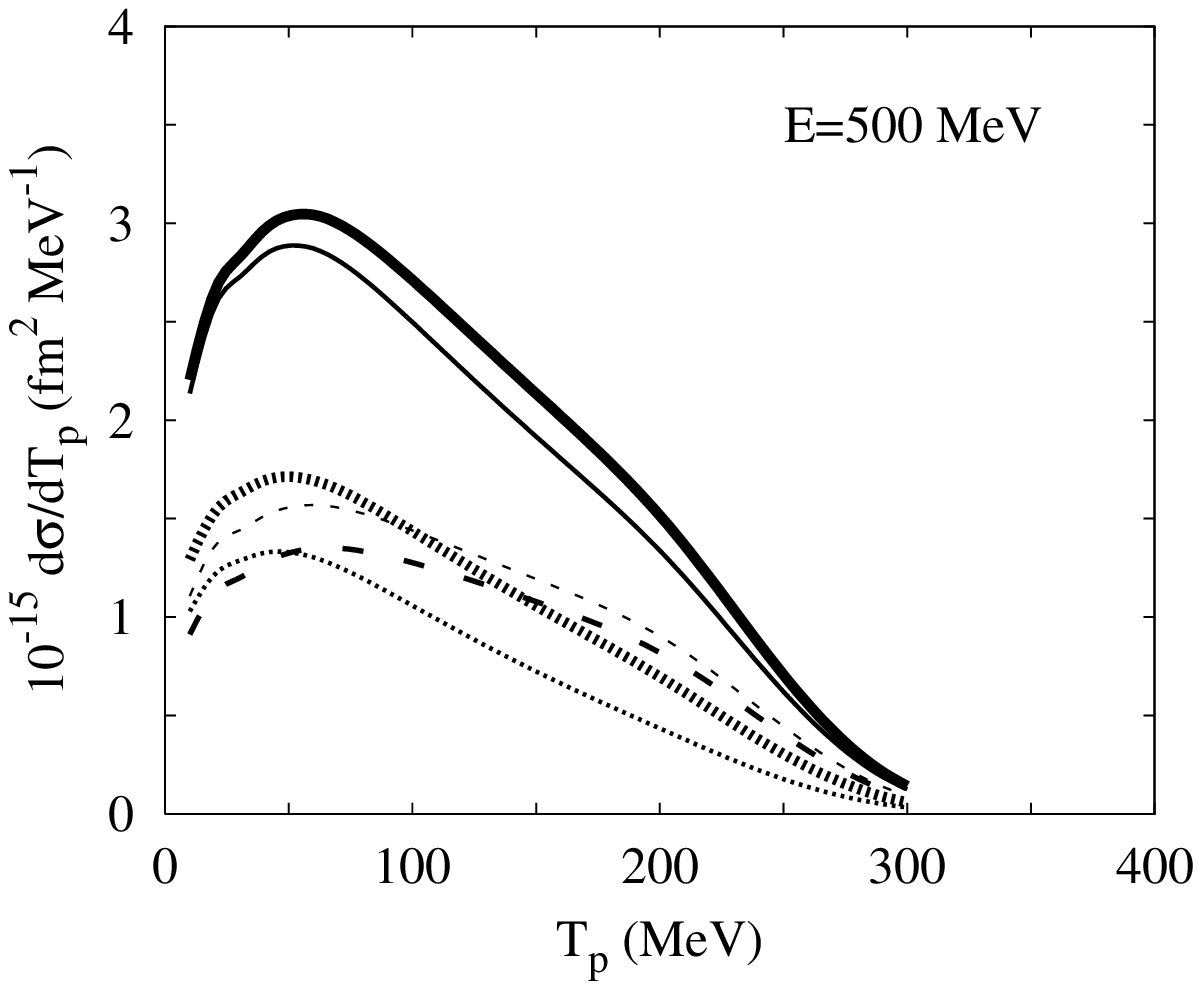}
\includegraphics[width=0.5\linewidth]{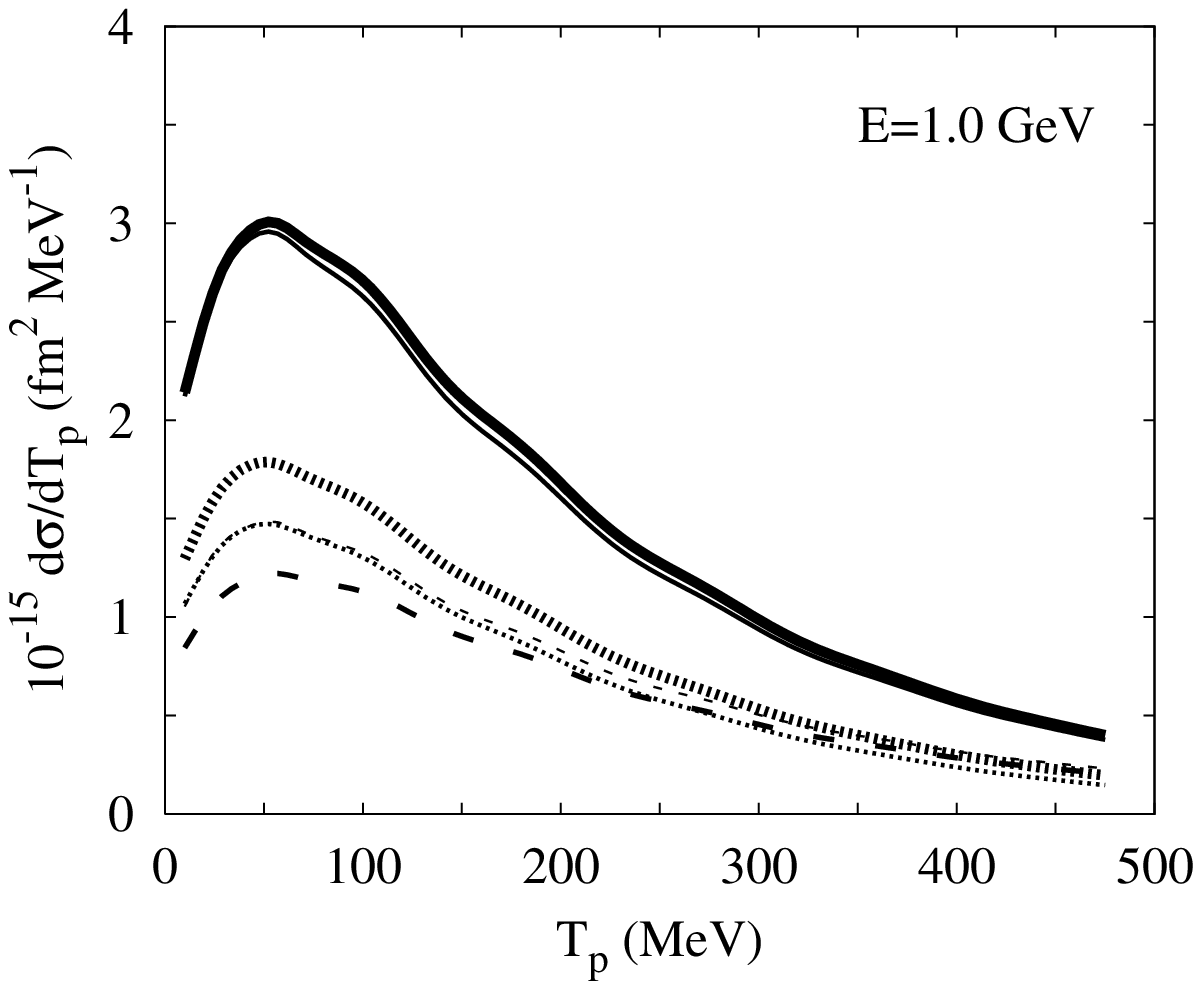}
\caption{Neutral current $^{12}$C($\nu, \nu'$) cross section as a
function of the knocked out nucleon kinetic energy $T_p$ at
incident neutrino energy $E=500$ MeV and 1.0 GeV. Solid curves are
the results for the cross sections, dashed and dotted lines are
the contributions of the proton and the neutron, respectively.
Thick (thin) lines are calculations without (with) the strange
quark contribution . The optical potential of the knocked-out
nucleon is used for the final state interaction.} \label{neut-gs}
\end{figure}

\begin{figure}
\includegraphics[width=0.5\linewidth]{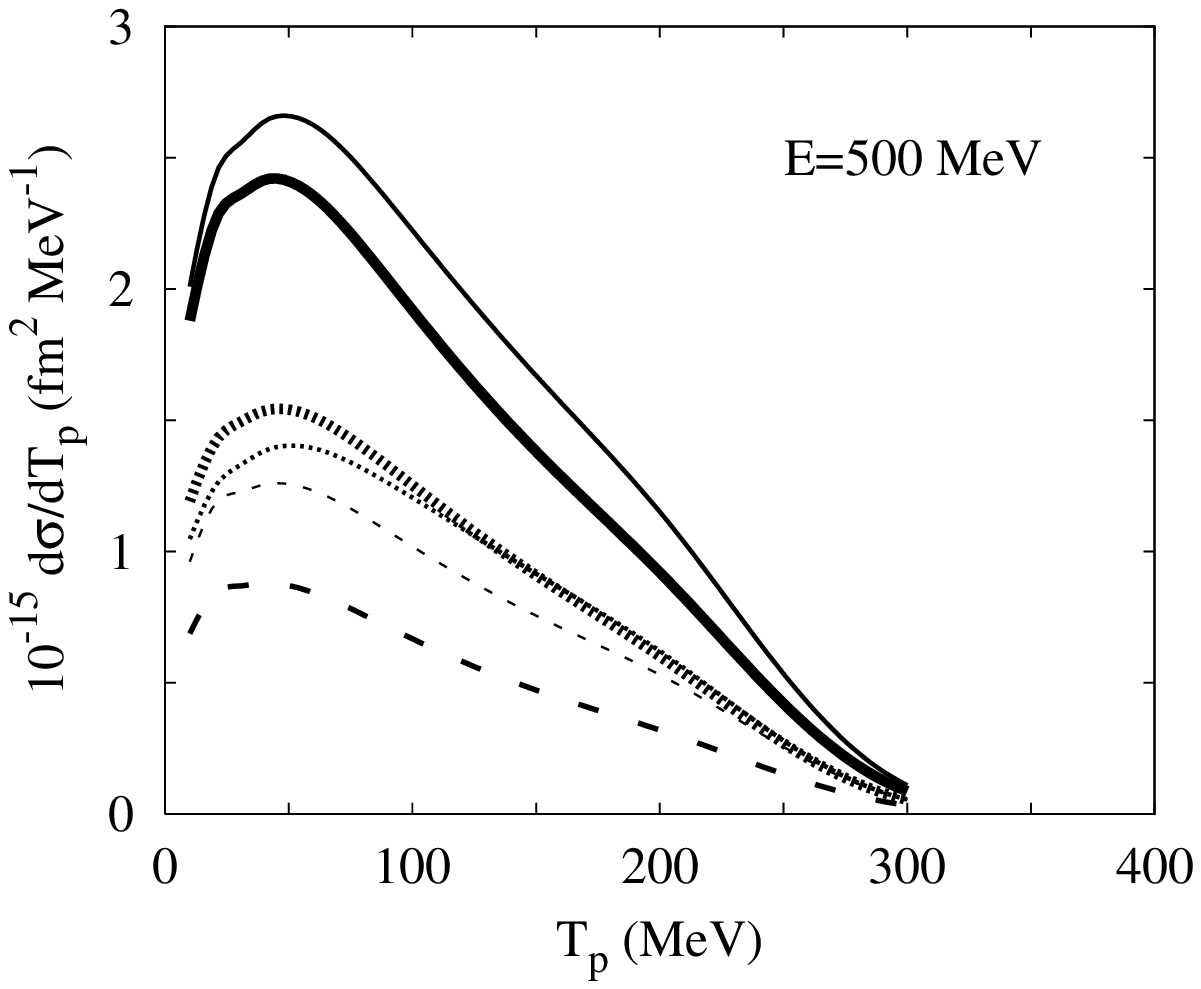}
\includegraphics[width=0.5\linewidth]{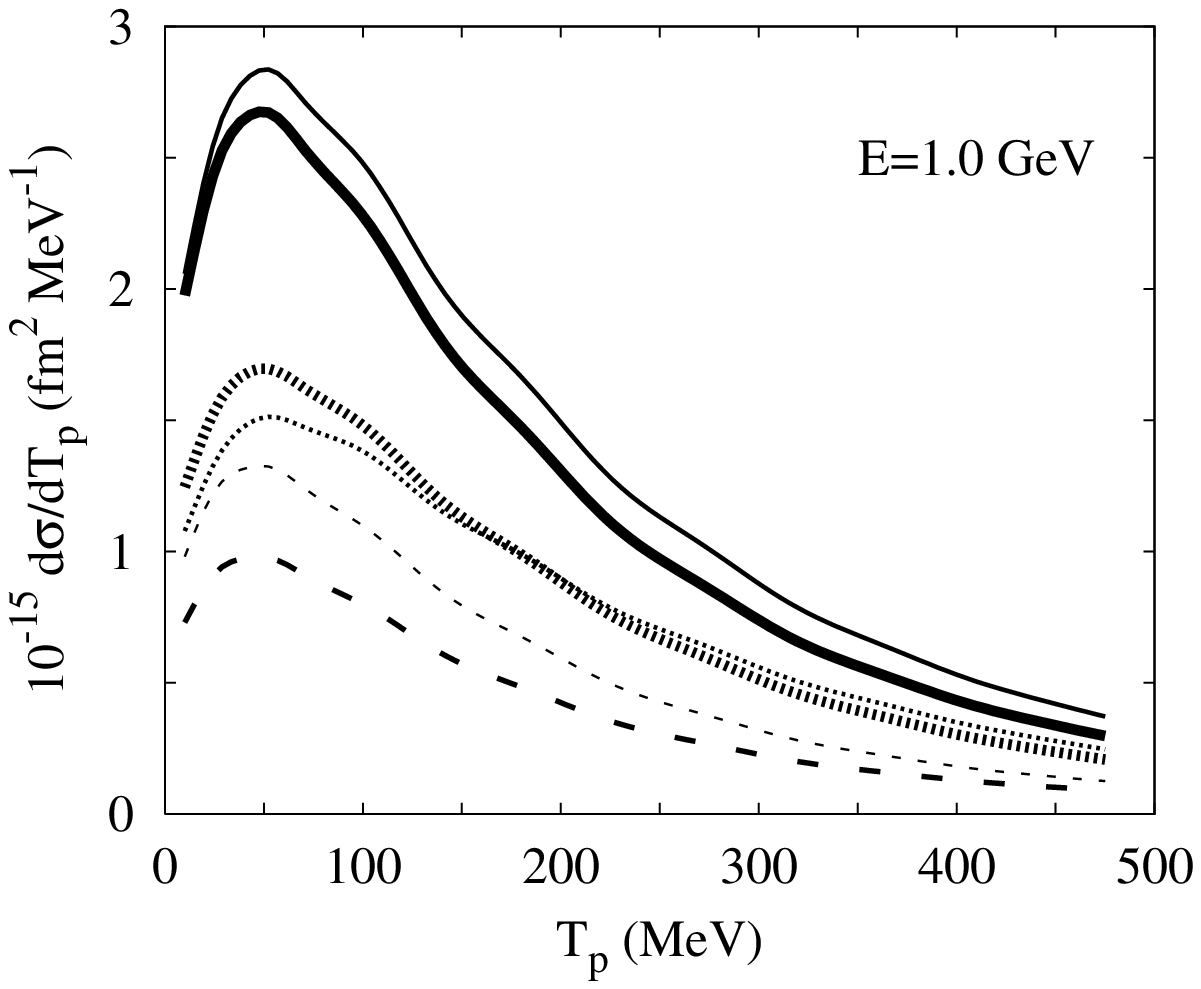}
\caption{The same as in Fig. \ref{neut-gs} but for the
antineutrino.} \label{anti-gs}
\end{figure}

Figures. \ref{neut-gs} and \ref{anti-gs} exhibit the effect of a
strange quark contribution to the axial form factor for the
neutrino and the antineutrino. The kinematics are the same as in
Fig. \ref{neut}. Thick (thin) curves are the results without
(with) the strange quark contribution. Note that the relativistic
optical potentials are used in these cases. For the incident
neutrino, the effect of the strange quark reduces the cross
section by the amount of 5\% for 500 MeV and 2\% for 1.0 GeV. The
effect of the strange quark for the knocked-out protons is
increased about 15\% for 500 MeV and 20\% for 1.0 GeV, but for the
corresponding knocked-out neutrons decreased about 30\% for 500
MeV and 20\% for 1.0 GeV. This different behavior is due to the
sign of $g_A$ in Eq. (\ref{gs}).

For the incident antineutrino, the effect of the strange quark
enhances the cross section by 10\% for 500 MeV and 5\% for 1.0
GeV, contrary to the neutrino case. For the knocked-out neutrons,
the effect constricts the results by 10\% for 500 MeV and by 12\%
for 1.0 GeV and enhances them by 40\% for 500 MeV and 35\% for 1.0
GeV for the protons. From these results, we learn that the effect
of the strange quark contribution changes the magnitude of the
cross section while the position of the peaks and the shape are
not affected. The effect of the strange quark contributes more to
the knocked-out protons than the neutrons in the corresponding
cross sections.

The intrinsic polarization of the neutrino and the antineutrino,
which is the difference between the neutrino and the antineutrino,
is reflected only on the third term in Eq. (\ref{cs}) due to its
helicity dependence. Using the helicities, the asymmetry for the
neutrino and the antineutrino is written as
\begin{equation}
A_l = {\frac {\sigma(h=-1) - \sigma(h=+1)} {\sigma(h=-1) +
\sigma(h=+1)}} , \label{asy}
\end{equation}
where $\sigma$ denotes the cross section in Eq. (\ref{cs}) and
$h=-1$ ($h=+1$) presents the helicity of the incident neutrino
(antineutrino).

In Fig. \ref{asy}, we study the effect of the FSI using the
asymmetry of the neutrino and the antineutrino at the incident
energies $E=500$ MeV and 1.0 GeV. Solid and dashed lines represent
the results with and without the FSI, respectively. The effects of
the FSI on the asymmetry appear to be much smaller than those on
the corresponding cross sections. The asymmetry itself does not
give any drastic effects coming from the FSI. But the effect
increases with lager kinetic energies of the knocked-out nucleon.

\begin{figure}
\includegraphics[width=0.5\linewidth]{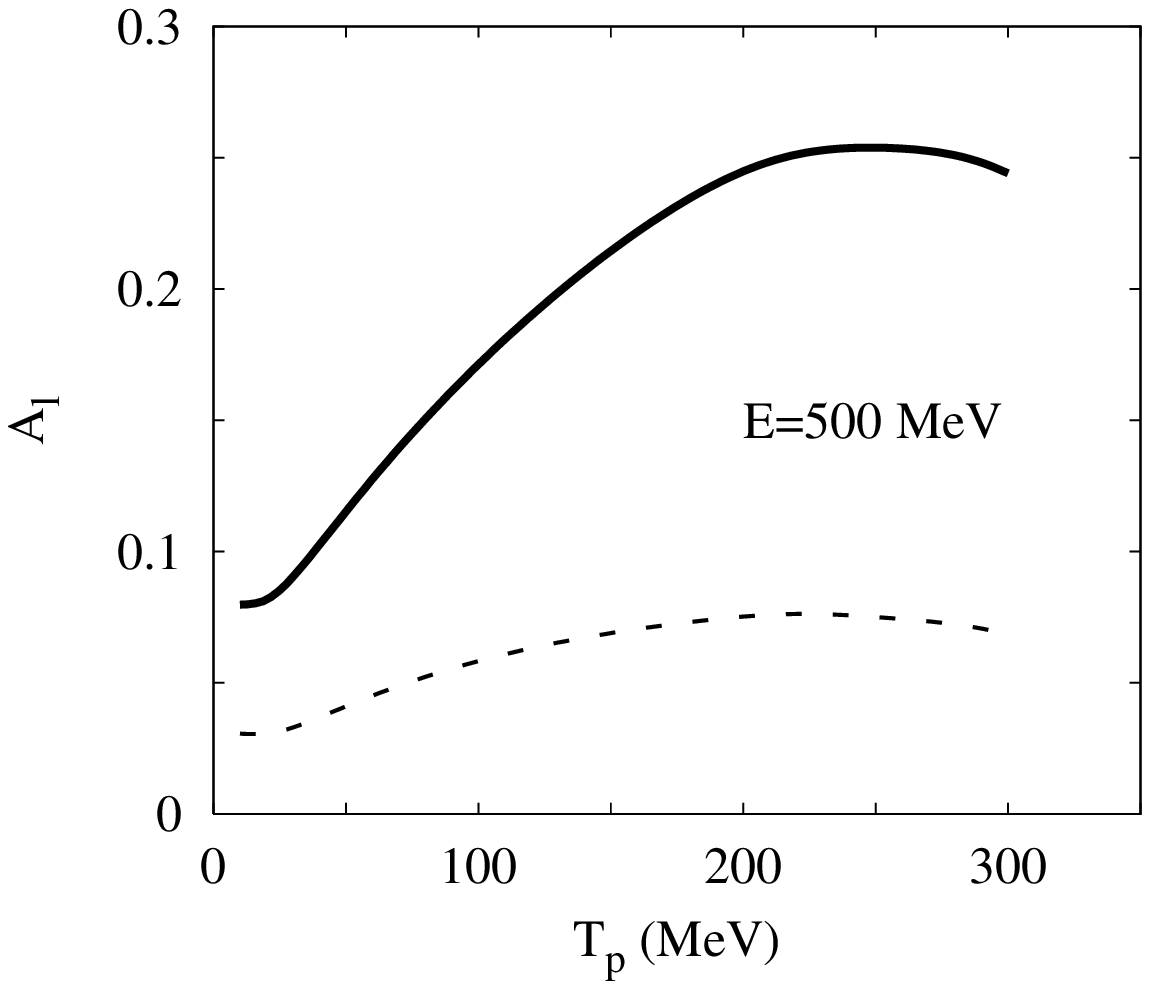}
\includegraphics[width=0.5\linewidth]{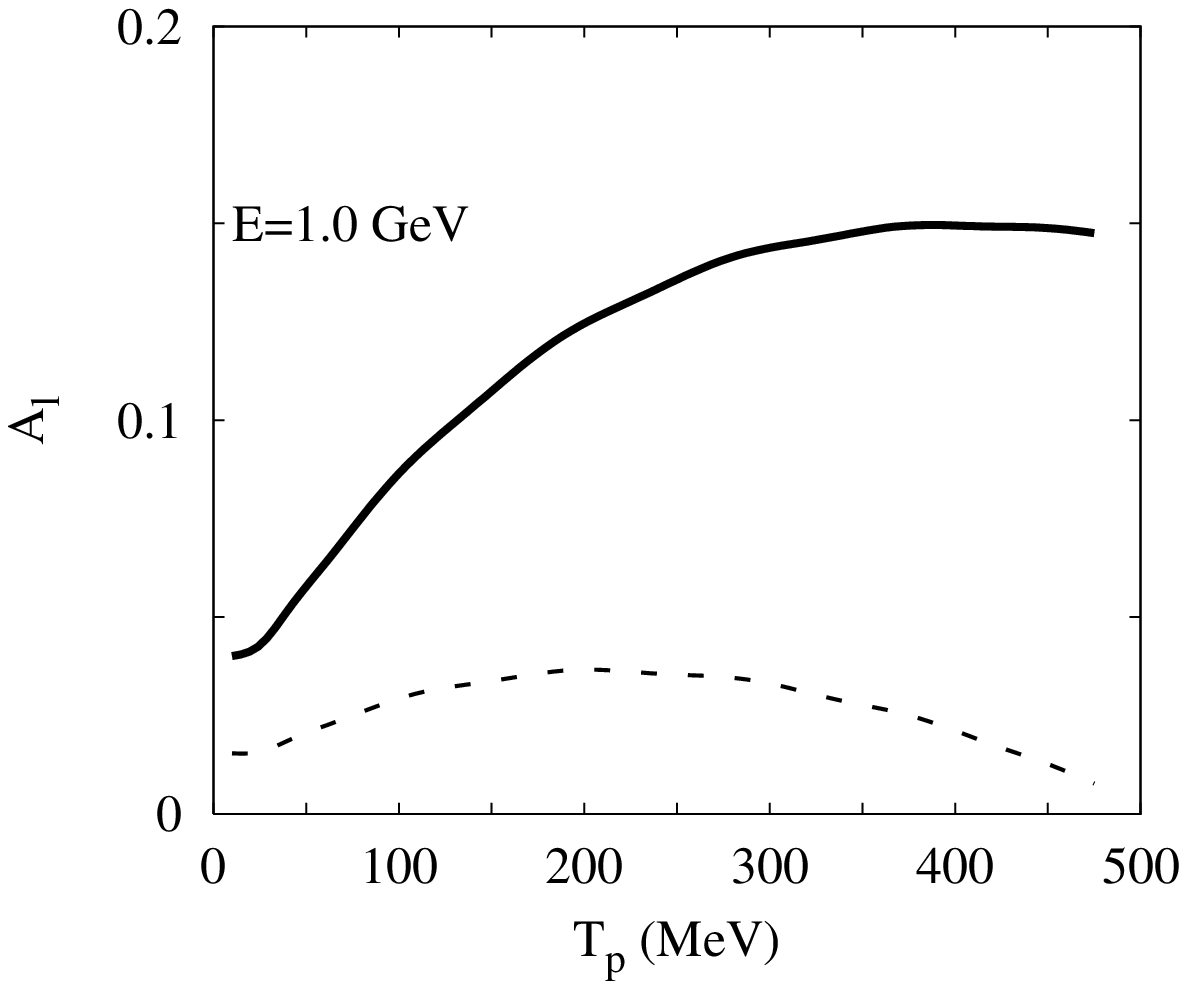}
\caption{The asymmetry as a function of the kinetics energy of the
knocked-out nucleon. Solid (dashed) curves represent the results
without (with) the strange quark contribution.} \label{asy-gs}
\end{figure}

With the same method in Fig. \ref{asy}, the effect of the strange
quark contribution is investigated in Fig. \ref{asy-gs}. Solid
(dashed) curves represent the results without (with) the strange
quark contribution. Surprisingly, the effect appears to be very
large comparing with the corresponding cross sections. The effect
also becomes larger as the kinetic energies of the knocked-out
nucleon increase. Therefore the asymmetry might be an effective
test for the strange quark contribution on the axial vector form
factor according to these calculations. Note that the FSI is
included through the optical potentials in these calculations.

\section{summary and conclusion}
We present the calculations of the neutral-current reaction for
the neutrino-nucleus scattering of $^{12}$C in quasielastic
region. The incident neutrino energies are used in intermediate
energies, 500 MeV and 1.0 GeV. The wave functions for bound state
are generated by the strong scalar and vector potentials based on
$\sigma - \omega$ model, and for the knocked-out nucleon by the
phenomenological optical model.

The FSI associated with the optical potential produces a large
reduction of the cross section and influences differently on the
ejecting proton and neutron. The effect appears to be larger on
the neutrino than on the antineutrino. On the other hand, by the
effect of the strange quark contribution, the cross sections are
reduced for the neutrino and enhanced for the antineutron. The
strange quark contribution shows a bigger contribution to the
reaction by the antineutrino than by the neutrino. For the
asymmetry of the intrinsic polarization of the neutrino and the
antineutrino, the effects of the FSI and the strange quark
increase with higher knocked-out nucleon kinetic energies.
According to Ref. \cite{giusti2}, the effect of the strange quark
is negligible small but in our calculation it is not small for the
asymmetry.

In conclusion, the effects of the final state interaction and the
strange quark contribution on the intrinsic polarization asymmetry
exhibit very different behavior on the corresponding cross
sections. It needs to investigate the effects on various structure
functions extracted from the methods commonly used in electron
scattering like $(e,e'p)$ and $(e,e')$ reactions. As one of next
works, it will be possible to include the Coulomb distortion of
the final leptons using charged-current neutrino-nucleus
scattering.

\newpage

\begin{figure}
\includegraphics[width=0.5\linewidth]{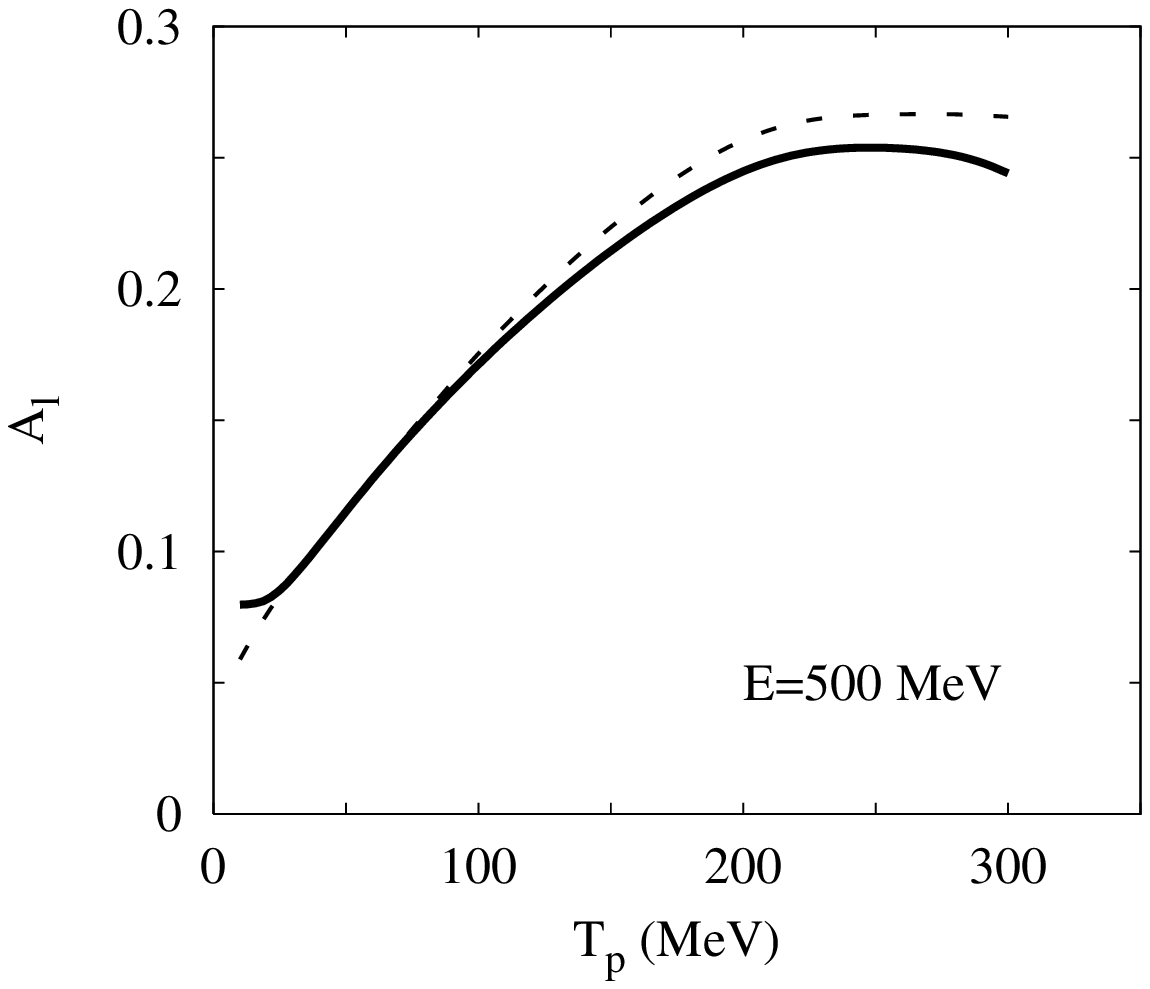}
\includegraphics[width=0.5\linewidth]{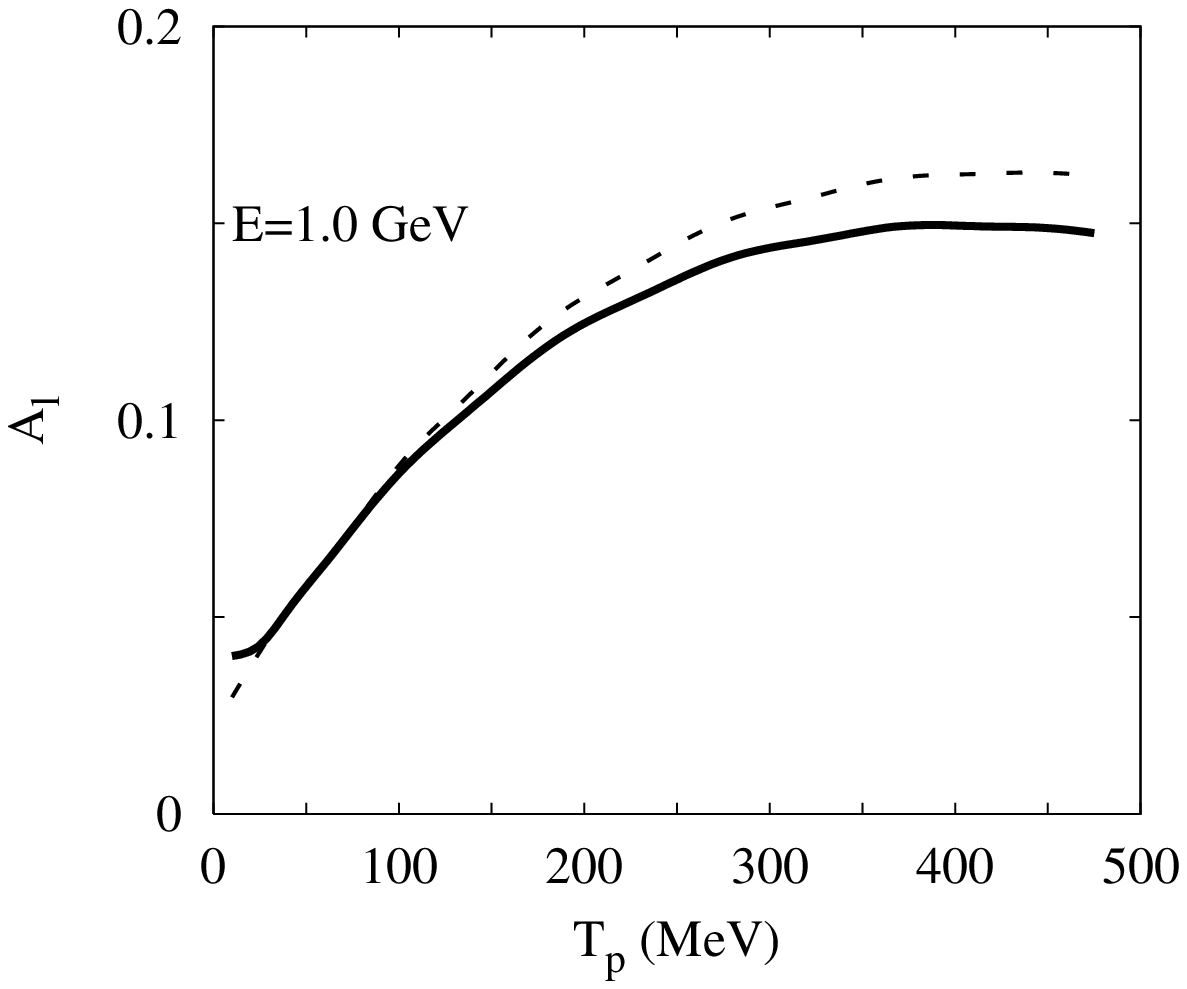}
\caption{The asymmetry as a function of the kinetics energy of the
knocked-out nucleon. Solid and dashed curves represent the results
with and without the final state interaction, respectively.}
\label{asy}
\end{figure}

\end{document}